\documentclass[aps,prb,reprint,superscriptaddress,amsmath,citeautoscript,flushbottom,floatfix]{revtex4-2}

\usepackage{graphicx}
\usepackage[colorlinks=true,linkcolor=black,citecolor=blue,urlcolor=blue]{hyperref}

\newcommand{\vc}[1]{\boldsymbol{#1}}
\newcommand{\pd}{{\phantom{\dagger}}}

\begin{document}

\title{Superconductivity in the spin-state crossover materials: 
Nickelates with planar-coordinated low-spin Ni$^{2+}$ ions} 

\author{Ji\v{r}\'{\i} Chaloupka}
\affiliation{Department of Condensed Matter Physics, Faculty of Science,
Masaryk University, Kotl\'a\v{r}sk\'a 2, 61137 Brno, Czech Republic}

\author{Giniyat Khaliullin}
\affiliation{Max Planck Institute for Solid State Research,
Heisenbergstrasse 1, D-70569 Stuttgart, Germany}

\begin{abstract}
We theoretically study quasi-two-dimensional nickel compounds, where the
nickel ions assume Ni$^{2+}$ $d^8$ valence state and feature a low-spin $S=0$
ground state quasidegenerate with $S=1$ ionic excitations. Such a level
structure is supported by square-planar coordination of nickel ions or a
suitable substitution of apical oxygens. We construct the corresponding
singlet-triplet exchange model and explore its phase diagram and excitation
spectrum. By hole doping, we further introduce mobile Ni$^{3+}$ $d^7$
ionic configurations with effective spins $S=1/2$, and analyze their
interactions with the $d^8$ singlet-triplet background. The interplay with
the triplet excitations in the $d^8$ sector is found to have a deep impact on
the propagation of the doped hole-like charge carriers and is identified as a
powerful source of Cooper pairing among them.
\end{abstract}

\date{\today}

\maketitle


\section{Introduction}

After decades of efforts to find cuprate-like superconductivity (SC) in
nickel-based compounds, unconventional SC has finally been discovered in a
number of nickelate families in recent
years~\cite{Li19,Sun23,Wan24,Ko25,Zho25,Zhu24,Zha25a,Zha25b}. It is truly
remarkable that the Ni valence states, orbital degeneracy, and fermiology vary
broadly among the nickelate superconductors: While \mbox{NdNiO$_2$} discovered
in 2019~\cite{Li19} apparently falls well into the cuprate's ``single-layer,
single-orbital, spin one-half'' paradigm, the bilayer
\mbox{La$_3$Ni$_2$O$_7$}~\cite{Sun23} and trilayer
\mbox{La$_4$Ni$_3$O$_{10}$}~\cite{Zhu24,Zha25a} nickelate SCs seem greatly
differ from cuprates in all respects. Indeed, \textit{(i)}~in contrast to only
weakly coupled CuO$_2$ planes in cuprates, the \mbox{NiO$_2$} planes are
tightly bound into bilayers or trilayers in them; \textit{(ii)}~distinct from
cuprates, both $e_g$ orbitals $x^2-y^2$ and $3z^2-r^2$ are essential; and,
finally, \textit{(iii)}~Ni-valence is far from the cuprate-like spin one-half
$d^9$ configuration, but rather in the mixed-valence regime with $\sim 7.5$
electrons in the $3d$ shell. Such a diversity of Ni-based SCs should be
related to the rich spin-and-orbital physics typical for nickelates, and
suggests that more superconductors may be found in the future among the
nickel-based materials with different lattice and orbital structures. 

\begin{figure}[tb]
\includegraphics[scale=1.0]{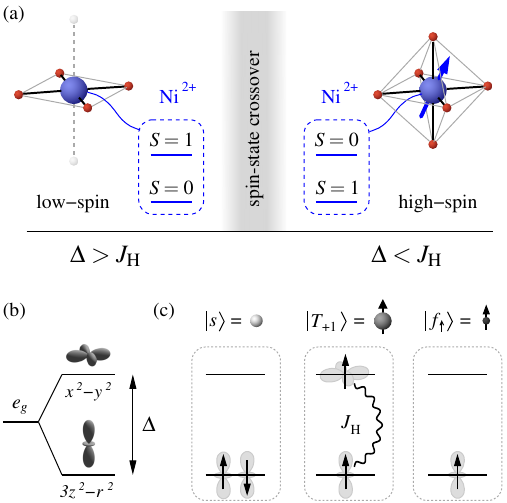}
\caption{
(a)~Ground-state spin configuration of a Ni$^{2+}$ ion with $d^8
(t_{2g}^6e_g^2)$ electronic filling, tuned by the competition of Hund's
coupling $J_\mathrm{H}$ and the crystal field splitting $\Delta$. If the
Ni$^{2+}$ ion is planar-coordinated or placed into a highly elongated oxygen
octahedra, it would assume a singlet ground state with doubly occupied
$3z^2-r^2$ orbital.
(b)~Splitting of $e_g$ orbital levels in a crystal field promoting the above
low-spin configuration.
(c)~Low-energy states selected as a basis of our effective model: $d^8$
singlet $s$ with doubly occupied $3z^2-r^2$ orbital, $d^8$ triplet $T$, where
the CF splitting $\Delta$ is nearly compensated by Hund's coupling
$J_\mathrm{H}$, and $d^7$ state corresponding to the hole-like particle $f$
with spin one-half. 
}\label{fig:ionic}
\end{figure}

One well-known consequence of the orbital degeneracy in compounds of
transition metal ions such as Fe, Co, and Ni are the famous spin-state
crossover phenomena~\cite{Gut04}. In spin-crossover materials, a close
competition between the lattice crystalline fields and intraionic Hund's
coupling results in the quasi-degeneracy of the ionic spin states. For
example, the Ni$^{2+}$ ion with $d^8$ configuration may adopt either $S=0$
singlet or $S=1$ triplet ground state, as illustrated in
Fig.~\ref{fig:ionic}(a). Importantly, the ground state spin value depends
sensitively on the lattice structure details, covalency, etc, and can thus be
tuned by external factors such as pressure or strain~\cite{Par12}. The
spin-state quasi-degeneracy is quite common phenomena among the Ni$^{2+}$
compounds with various lattice structures, e.g.,
\mbox{Sr$_2$NiO$_2$Cu$_2$(S,Se)$_2$}~\cite{Smy22,Otz99,Cla08},
\mbox{(Sr,Ba)$_2$NiO$_2$Ag$_2$Se$_2$}~\cite{Mat19},
\mbox{BaNiO$_2$}~\cite{Mat99}, \mbox{LaNiO$_{2.5}$}~\cite{Alo97}. 

In this paper we perform a model exploration of single-layer nickelates in
the spin-state crossover regime. Specifically, we consider low-spin Ni$^{2+}$
$d^8$ Mott insulators doped by mobile hole-like charge carriers, and address the
question of pairing mechanisms and possible superconductivity in them. We find
that the doped holes strongly interact with the spin-state fluctuations,
namely, while moving they emit and absorb singlet-triplet excitations. This
leads to polaron physics similar to that in the electron-phonon problem, and
gives rise to an effective attraction between the charge carriers. We analyze
and evaluate various pairing processes, and conclude that the nickel-based
spin-crossover materials are promising candidates to host high-temperature
superconductivity.

The paper is organized as follows: Section~\ref{sec:d8} derives the exchange
model for the undoped Mott insulator with low-spin Ni$^{2+}$ $d^8$ ions
(Sec.~\ref{sec:d8model}), and presents its phase diagram along with a
cross-check against exact diagonalization of the underlying Hubbard model
(Sec.~\ref{sec:d8phases}). The mobile $d^7$ hole-like carriers are considered
in Sec.~\ref{sec:d7}: we first introduce the interactions between the holes
and the $d^8$ background (Sec.~\ref{sec:d7model}), then analyze the
single-particle propagation of the doped holes at low densities
(Sec.~\ref{sec:d7SCBA}), and finally move on to the central topic of the
paper---a study of the mechanisms of Cooper pairing among the holes
(Sec.~\ref{sec:d7pairing}). Section~\ref{sec:Conclusions} concludes the
paper. 


\section{Low-spin $\mathbf{Ni}^\mathbf{2+}$ $d^8$ insulator}
\label{sec:d8}

We construct our model as a low-energy effective model of the Hubbard model
based on $e_g$ orbitals. We assume that crystal field splitting $\Delta$
between $x^2-y^2$ and $3z^2-r^2$ orbitals [Fig.~\ref{fig:ionic}(b)] is
sufficient to compensate the Hund's coupling $J_\mathrm{H}$ and stabilize
spin-singlet ground state. It is important to note at this point that
$J_\mathrm{H}$ in a solid is well reduced from its free-ion value due to
covalency effects. 

We further assume that the balance between $\Delta$ and $J_\mathrm{H}$ is such
that while the ground state of Ni$^{2+}$ is nonmagnetic, the excited
spin-triplet states are still nearby in energy. In this regime, the relevant
basis states are those depicted in Fig.~\ref{fig:ionic}(c). As a first step,
we consider the non-doped case with $d^8$ ($t_{2g}^6e_g^2$) electronic
configuration, and study the corresponding exchange model based on the ionic
singlet $s$ and triplet $T$ states.

\subsection{Exchange model}
\label{sec:d8model}

Built on top of a $t^6_{2g}$ configuration, which is considered rigid, the
basis states $s$ and $T$ are obtained by diagonalizing the on-site Coulomb
interaction between the two additional $e_g$ electrons, as embedded in the
local part of the $e_g$ Hubbard model:
\begin{multline}
\tfrac12\Delta\, (n_x-n_z)
+ U (n_{x\uparrow}n_{x\downarrow} + n_{z\uparrow}n_{z\downarrow}) 
+ (U-\tfrac52 J_\mathrm{H}) n_x n_z \\
- 2J_\mathrm{H} \vc S_x \vc S_z 
+ J_\mathrm{H} (
x^\dagger_{\uparrow}x^\dagger_{\downarrow} z^\pd_{\downarrow}z^\pd_{\uparrow} 
+ z^\dagger_{\uparrow}z^\dagger_{\downarrow} x^\pd_{\downarrow}x^\pd_{\uparrow} 
) \,.
\end{multline}
Here we have introduced a shorthand notation by denoting the electron
operators $d_{x^2-y^2}$ and $d_{3z^2-r^2}$ as $x$ and $z$, respectively. 
Due to the pair hopping term $\propto J_\mathrm{H}$, the singlet ground state
$|s\rangle$ is actually a linear combination
\begin{equation}
|s\rangle = \bigl( \cos\theta\; z^\dagger_\uparrow z^\dagger_\downarrow
                  -\sin\theta\; x^\dagger_\uparrow x^\dagger_\downarrow \bigr) 
\;|t_{2g}^6\rangle
\end{equation}
with $\theta$ given by $\tan 2\theta = J_\mathrm{H}/\Delta$. In the regime of
our interest, $\theta$ is small and most of the weight is carried by the pair
of electrons in $3z^2-r^2$ orbital as indicated in Fig.~\ref{fig:ionic}(c). 
The three members of the triplet $T$ read simply as
\begin{align}
|T_{+1}\rangle &= x^\dagger_\uparrow z^\dagger_\uparrow
\;|t_{2g}^6\rangle \,, \notag \\
|T_{0}\rangle &= \tfrac{1}{\sqrt{2}}\bigl(
 x^\dagger_\uparrow z^\dagger_\downarrow +
 x^\dagger_\downarrow z^\dagger_\uparrow \bigr)
\;|t_{2g}^6\rangle \,, \notag \\
|T_{-1}\rangle &= x^\dagger_\downarrow z^\dagger_\downarrow
\;|t_{2g}^6\rangle \,. \label{eq:Tstates}
\end{align}
The energy splitting between the above states is an essential parameter of our
model. Based solely on the ionic Hamiltonian, we get for the singlet-triplet
splitting
\begin{equation}\label{eq:ET}
E_T=E(|T\rangle)-E(|s\rangle) = 
\sqrt{\Delta^2+J_\mathrm{H}^2}-3J_\mathrm{H} \,.
\end{equation}
In reality, this splitting is influenced by various factors beyond the ionic
Hamiltonian, for example covalency effects, and may be therefore regarded as a
free parameter. However, for simplicity we will maintain the connection to the
$e_g$ Hubbard model and use Eq.~\eqref{eq:ET} hereafter. Note, that if the
pair hopping term is omitted, we get the familiar expression
$\Delta-3J_\mathrm{H}$ for $E_T$.

The exchange interactions among $d^8$ ions are derived in the usual way by
perturbatively eliminating the intersite hopping to second order. In the case
of $e_g$ orbitals on the square lattice, the hopping at nearest-neighbor bond
$ij$ takes the following form fixed by Slater-Koster rules:
\begin{equation}\label{eq:hopeg}
-t \,\left[ x^\dagger x + \tfrac13 z^\dagger z
\mp\tfrac1{\sqrt3}(x^\dagger z + z^\dagger x)
\right]_{ij} \,.
\end{equation}
We parametrize the hopping by the larger amplitude $t$ for the planar
$x^2-y^2$ orbitals. The doped holes to be introduced later will reside in
$3z^2-r^2$ orbitals, having thus a significantly reduced bandwidth determined
by $t/3$. The sign $(\mp)$ of the interorbital hopping is linked to the bond
direction: $-$ is applied at $x$ bonds, $+$ at $y$ bonds.

\begin{figure}[tb]
\includegraphics[scale=1.0]{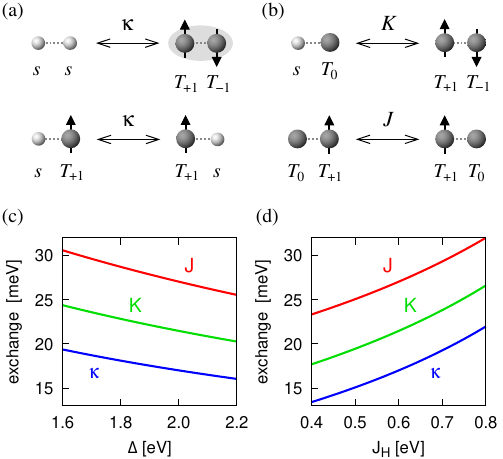}
\caption{
(a)~Examples of exchange processes among the $d^8$ ions contributing to the
$\kappa$ interaction channel of $\mathcal{H}_{d^8}$ of Eq.~\eqref{eq:Hd8}.
There are two distinct classes of processes---creation/annihilation of
singlet pairs of triplons $\vc T$ (top) and $\vc T$ hopping in the $s$
background (bottom).
(b)~Sample processes contained in the higher-order channels $K$ and $J$ of
$\mathcal{H}_{d^8}$.
(c)~Exchange parameters entering Eq.~\eqref{eq:Hd8} as obtained by
second-order perturbation theory [see Eq.~\eqref{eq:exchpar}] assuming
$U=5\:\mathrm{eV}$, $J_\mathrm{H}=0.6\:\mathrm{eV}$, and $t=0.3\:\mathrm{eV}$.
(d)~The same for fixed $\Delta=2\:\mathrm{eV}$ and varying $J_\mathrm{H}$.
}\label{fig:d8model}
\end{figure}

The exchange interactions are most intuitively depicted as bond processes
involving $s$ (scalar) and $\vc T$ (vector) hardcore bosons representing the
singlet and triplet $d^8$ states. The hardcore bosons $\vc T$ are hereafter
referred to as \textit{triplons}. Such a hardcore-boson representation is used
in Fig.~\ref{fig:d8model}(a),(b), where we show a few examples of the rather
rich possibilities to be discussed later. Fortunately, due to the inherent
spin-isotropy imposed by the underlying Hubbard model, the exchange
interactions have to obey spin conservation rules, which limits the number of
independent model parameters. To express the resulting $d^8$ model
Hamiltonian in a compact form, we utilize several auxiliary vector operators
based on the following Cartesian combinations of the triplet states:
\begin{align}
|T_x\rangle &= \tfrac{i}{\sqrt{2}}(|T_{+1}\rangle-|T_{-1}\rangle) \,, \notag \\
|T_y\rangle &= \tfrac{1}{\sqrt{2}}(|T_{+1}\rangle+|T_{-1}\rangle) \,, \notag \\
|T_z\rangle &= -i|T_0\rangle \,. \label{eq:TCart}
\end{align}
The first vector operator $\vc{\widetilde{S}}=
(\widetilde{S}^x,\widetilde{S}^y,\widetilde{S}^z)$ is local and 
is associated with the on-site $s\leftrightarrow T$ transition:
\begin{equation}\label{eq:Stildedef}
\widetilde{S}^\alpha = -i\,(s^\dagger T^\pd_\alpha - T^\dagger_\alpha s)
\,,\quad \alpha=x,y,z \,.
\end{equation}
The second local operator $\vc S=(S^x,S^y,S^z)$ directly corresponds to the
spin-1 carried by $\vc T$ and is defined by
\begin{equation}\label{eq:Sdef}
S^\alpha = -i\,\epsilon_{\alpha\beta\gamma} T^\dagger_\beta T^\pd_\gamma
\,,\quad\text{i.e.}\quad 
\vc S=-i(\vc{T}^\dagger\times\vc{T})
\,.
\end{equation}
We also introduce analogous bond operators having the internal structure
identical to $\vc{\widetilde{S}}$ and $\vc S$:
\begin{equation}\label{eq:Sbonddef}
\widetilde{S}^\alpha_{ji} = -i\,(s^\dagger_j T^\pd_{\alpha i} 
- T^\dagger_{\alpha j} s^\pd_i)
\,,\quad
S^\alpha_{ji} = -i\,\epsilon_{\alpha\beta\gamma}
T^\dagger_{\beta j} T^\pd_{\gamma i} 
\,.
\end{equation}
The above operators enable us to express the $d^8$ model Hamiltonian in a
manifestly spin-isotropic compact form
\begin{multline}\label{eq:Hd8}
\mathcal{H}_{d^8} = E_T \sum_i n_{Ti}
+\kappa \sum_{\langle ij \rangle} 
\vc{\widetilde{S}}_i\,\vc{\widetilde{S}}_j \\
+\sum_{\langle ij \rangle} 
\left[
\mathrm{sgn}_{ij}\, 
K( \vc{\widetilde{S}}_{ij}\vc{S}_{ji}
  +\vc{S}_{ij}\vc{\widetilde{S}}_{ji} )
+ J\,\vc{S}_i\,\vc{S}_j 
\right] \,.
\end{multline}
Here the first term counts the number of triplons via 
$n_T=\sum_{\alpha=x,y,z} T^\dagger_\alpha T^\pd_\alpha$ 
and penalizes them by the energy $E_T$, the other terms represent the exchange
interactions being sorted as second, third, and fourth-order contributions in $T$
operators, i.e. according to their expected relevance at low triplon density.
Due to $e_g$ orbital symmetry, there is sensitivity to the bond direction
inherited from the hopping \eqref{eq:hopeg}. It is captured by the
bond-dependent sign factor $\mathrm{sgn}_{ij}$, which is equal to $+1$ or $-1$
for $x$ and $y$ bonds, respectively. Finally, when rewriting the operators in
the $K$ term using the elementary hardcore-boson operators $s$ and $\vc{T}$,
these have to be taken in normal order with the creation operators moved to
the left.

We now briefly discuss the rather complex exchange processes contained in
\eqref{eq:Hd8} with the help of the cartoon representations in
Fig.~\ref{fig:d8model}(a),(b), capturing the exchange in the form of a
generation, propagation, and mutual interactions of hardcore triplet particles
$\vc T$ in a background made of $s$. The $\kappa$-term
$\vc{\widetilde{S}}_i\,\vc{\widetilde{S}}_j$ includes two distinct types of
processes depicted in Fig.~\ref{fig:d8model}(a). The first is a creation and
annihilation of singlet pairs of $\vc T$ on bonds, corresponding to a
combination
$T_xT_x+T_yT_y+T_zT_z = T_{+1}T_{-1}-T_{0}T_{0}+T_{-1}T_{+1}$.
The other one is a simple $\vc T$ hopping preserving the $T$ label (either
$\pm 1$, $0$ or Cartesian $x,y,z$). The relative amplitude of these processes
is fixed to $-1$ by the interaction term. Being of second order in $T$
operators, this is the most important exchange interaction at low triplon
concentration of about $n_T \lesssim 0.2$ per site. The third-order $K$
contribution may be understood as an $s\leftrightarrow T$ transition entangled
with the spin-1 of another $\vc T$ particle, while the fourth-order $J$ term
is just the regular Heisenberg exchange between the spins-1 carried by $\vc
T$. Two examples of such higher-order processes are presented in
Fig.~\ref{fig:d8model}(b).

Through the perturbative calculation, one arrives at two more contributions
not explicitly included in \eqref{eq:Hd8}: \textit{(i)}~a small
renormalization of $E_T$ level, \textit{(ii)}~repulsion between $T$ particles.
Both are marginal, opposing each other, and can be safely absorbed into $E_T$
(the latter one on a mean-field level), so that they do not alter the form of
the $d^8$ model \eqref{eq:Hd8}.

The overall phase behavior of the $d^8$ model \eqref{eq:Hd8} is driven by the
competition between the triplon cost $E_T$ and the strength of the exchange
interactions that profit from the presence of the triplons and eventually lead
to their condensation. The microscopic derivation provides the following
values of the three exchange parameters:
\begin{align}
\kappa &= \frac{t^2}{3} \left(
\frac{1+\cos2\theta}{E_{-1}}
+\frac{2\sin2\theta}{E_{0}}
+\frac{1-\cos2\theta}{E_{+1}}
\right) 
, \notag \\
K &= \frac{2\sqrt2 \,t^2}{3\sqrt3} \left[ 
\left( \frac{1}{E_{-1}} + \frac{1}{E_{0}} \right)\cos\theta +
\left( \frac{1}{E_{+1}} + \frac{1}{E_{0}} \right)\sin\theta
\right]
, \notag \\
J &= \frac{t^2}{3} \left(
\frac{1}{E_{-1}} + \frac{10}{3E_{0}} + \frac{1}{E_{+1}}
\right) , \label{eq:exchpar}
\end{align}
where the denominators contain virtual energies 
$E_n=U+n\Delta+J_\mathrm{H}+2E_T$ ($n=0,\pm 1$) in the intermediate states.
Plotted in Fig.~\ref{fig:d8model}(c),(d) are the exchange constants evaluated
for representative values of the microscopic parameters. For the Ni ions, we
assume $U=5\:\mathrm{eV}$ and $J_\mathrm{H}=0.6\:\mathrm{eV}$ throughout the
paper, the crystal-field splitting $\Delta$ is varied around $2\:\mathrm{eV}$,
leading to quasidegenerate $s$ and $T$ levels. The hopping $t$ is set to a
representative value of $0.3\:\mathrm{eV}$ in Fig.~\ref{fig:d8model}. The
values of $\kappa$, $K$, and $J$ observed in Fig.~\ref{fig:d8model}(c),(d) are
relatively robust to variations in $\Delta$ and $J_\mathrm{H}$. This is
caused by the dominating values of $U+n\Delta$ in the denominators in
\eqref{eq:exchpar}, which are only little affected by changes $\Delta$ and
$J_\mathrm{H}$ within the ranges shown. In the presented parameter regime we
get the hierarchy $J>K>\kappa$, which, however, needs to be considered
together with the different role of these interactions. Even though $J$ and
$K$ are larger than $\kappa$, they are of higher order in $T$ operators and
thus relevant only at larger concentration $n_T$ of triplons. In addition, to
activate $J$ and $K$ interactions, a prior presence of triplons is necessary,
while $\kappa$ interaction channel generates them itself. In some of the cases
discussed later, it will be thus sufficient to study a simplified model
including $E_T$ and $\kappa$ terms only.


\subsection{Phase diagram and excitations}
\label{sec:d8phases}

A basic exploration of the phase behavior of the singlet-triplet model and its
excitation spectrum can be performed using the methods developed in the
context of spin ladders or bilayer magnets, where the singlet-triplet basis is
hosted by Heisenberg rungs connecting the rails or layers
\cite{Sac90,Gop94,Chu95,Som01}. The possibility of triplon condensation at a
sufficient exchange strength is addressed using the variational Ansatz
\begin{equation}\label{eq:trial}
|\Psi\rangle = \prod_{\vc R} \left[
\sqrt{1-\rho}\; s^\dagger + \sqrt{\rho}\, (\vc u-i\vc v)\, \vc T^\dagger
\right]_{\vc R} |\mathrm{vac}\rangle \,.
\end{equation}
The energy obtained by averaging the Hamiltonian \eqref{eq:Hd8} in the trial
state
\eqref{eq:trial}
\begin{multline}\label{eq:trialEavg}
E_\mathrm{avg} = E_T \sum_i \rho + \sum_{\langle ij\rangle}
\Bigl\{ 4\kappa \rho\,(1-\rho)\, \vc{v}_i\vc{v}_j \\
-\mathrm{sgn}_{ij}\, K \rho\sqrt{\rho\,(1-\rho)}
\;[\vc{v}_i(\vc u\times\vc v)_j+(\vc u\times\vc v)_i\vc{v}_j] \\
+4J\rho^2\, (\vc u\times\vc v)_i (\vc u\times\vc v)_j \Bigr\} 
\end{multline}
is to be minimized with respect to the condensate density $0\leq \rho\leq 1$
and site-dependent real vectors $\vc u$, $\vc v$ that obey the condition
$\vc{u}^2+\vc{v}^2=1$. This approach does not include quantum fluctuations
but still gives a semi-quantitatively correct picture, as we will demonstrate
later by cross checking with the exact diagonalization of the underlying
Hubbard model.

\begin{figure}[t!b]
\includegraphics[scale=1.0]{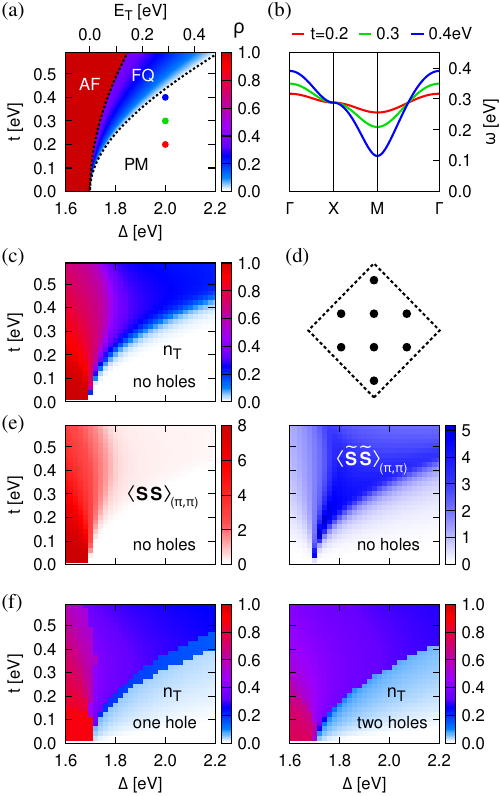}
\caption{
(a)~Phase diagram of the $d^8$ model in Eq.~\eqref{eq:Hd8} obtained by a
variational estimate based on the trial state \eqref{eq:trial}. Shown is the
condensate density $\rho$ which equals zero in the PM phase, saturates around
$0.3$ in the FQ phase and jumps to $\approx 1$ upon entering the AF phase.
The model parameters are given by Eqs.~\eqref{eq:ET} and \eqref{eq:exchpar}
evaluated for $U=5\:\mathrm{eV}$, $J_\mathrm{H}=0.6\:\mathrm{eV}$ and variable
$\Delta$ and $t$.
(b)~Triplon $\vc T$ dispersion in PM phase calculated for the three parameter
points marked in (a) ($\Delta=2\:\mathrm{eV}$ and $t=0.2$, $0.3$ or
$0.4\:\mathrm{eV}$) and plotted along the path connecting the high-symmetry
points $\Gamma=(0,0)$, $X=(\pi,0)$, and $M=(\pi,\pi)$ in the Brillouin zone.
At the PM/FQ boundary, the dispersion becomes gapless at the $M$ point.
(c)~Triplon number per site $n_T$ obtained by exact diagonalization of the
$e_g$ Hubbard model for the 8-site cluster shown in (d).
(e)~Characteristic correlations in the cluster ground state revealing the
tendencies towards long-range ordering---either AF (left) or FQ (right).
(f)~The same as in (c) but calculated for a reduced number of electrons on the
cluster to effectively create 1 or 2 holes.
}\label{fig:PD}
\end{figure}

Presented in Fig.~\ref{fig:PD}(a) is the variational phase diagram obtained by
varying the microscopic parameters $\Delta$ and $t$ and evaluating the
effective model parameters $E_T$, $\kappa$, $K$, $J$ via Eqs.~\eqref{eq:ET}
and \eqref{eq:exchpar}. The variational approach identifies three phases
described in the following.

In the paramagnetic (PM) phase, the condensate is absent ($\rho=0$) and
triplons $\vc T$ correspond to gapped elementary excitations. Their dispersion
can be obtained by starting with $\mathcal{H}_{d^8}$ of Eq.~\eqref{eq:Hd8},
making the replacement $s,s^\dagger\rightarrow\sqrt{1-n_T}$ to account for the
hardcore constraint, and expanding the result in $T_\alpha$,
$T_\alpha^\dagger$ operators. Afterwards, $\vc T$ are treated as
unconstrained bosons. At the same level of approximation as used when
constructing the phase diagram itself, the expansion is performed to second
order and the approximate Hamiltonian is readily solved by Fourier and
Bogoliubov transformations. This standard procedure \cite{Chu95} gives a
three-fold degenerate dispersion
\begin{equation}\label{eq:wqdisp}
\omega_{\vc q} = \sqrt{E_T(E_T+8\kappa\gamma_{\vc q})}
\end{equation}
with $\gamma_{\vc q}=\tfrac12(\cos q_x+\cos q_y)$, visualized in
Fig.~\ref{fig:PD}(b) for a few sample points in the PM phase. As the exchange
strength increases, following the increase of $t$, the excitations gradually
soften at the momentum $(\pi,\pi)$. When the critical point $E_T=8\kappa$ is
reached, the excitation gap closes, signaling a condensation of triplons and
the emergence of long-range order. 
Note that according to Eq.~\eqref{eq:wqdisp}, the ratio $\kappa/E_T$
determines the relative bandwidth of the triplon dispersion, while $E_T$ sets
the overall level. Therefore, if we approached the PM/FQ phase boundary in
Fig.~\ref{fig:PD}(a) by reducing $\Delta$ (and thus $E_T$) instead of
increasing $t$, the softening would proceed mainly as a downshift of the
dispersion, again touching zero at the critical point.
The observation of signatures of the above excitations, e.g. by neutron
scattering or resonant inelastic x-ray scattering, would be an important
experimental test of our model and would provide a valuable input for its
quantification in a particular material. We also note that a significant
broadening of the triplons due to their mutual interactions as well as
interactions with doped carriers can be expected. 

In the new phase, labeled by FQ in Fig.~\ref{fig:PD}(a), the optimization of
the variational Ansatz \eqref{eq:trial} leads to a nonzero condensate density
$\rho$. The condensate structure is given by staggered $\vc v_{\vc R} = 
\vc v \,\mathrm{e}^{i\vc Q\vc R}$ with $\vc Q=(\pi,\pi)$ and zero $\vc u$.
The corresponding energy per site derived from \eqref{eq:trialEavg} takes the
value $E_\mathrm{FQ} = -2\kappa(1-E_T/8\kappa)^2$ and does not include $K$ and
$J$ as these contribute at higher order via quantum fluctuations.
Based on the above condensate structure, the phase is characterized by
long-range correlations of the auxiliary quantity $\vc{\widetilde{S}}\propto
\vc v$ with the characteristic momentum $\vc Q$ and nonzero average
$\langle\vc{\widetilde{S}}\rangle_{\vc q=\vc Q}$ acting as an order parameter. 
In terms of the measurable spin-1 carried by $\vc T$, one gets the average
$\langle S^\alpha S^\beta \rangle_{\vc q=0} = \rho\,
(\delta_{\alpha\beta}-v_\alpha v_\beta)$ corresponding to a ferroquadrupolar
(FQ) order, where $\vc v$ plays the role of the director. Note, however, that
the primary order parameter is the above
$\langle\vc{\widetilde{S}}\rangle_{\vc q=\vc Q}$ and the FQ spin-1
correlations are reduced by having $\vc T$ mixed with still prevailing $s$.

Finally, if the triplon cost $E_T$ is low enough, the system switches into a
fully saturated antiferromagnetic (AF) phase with $\rho=1$, i.e. each site is
occupied by spin-1 $\vc T$ particle. This transition may be thus interpreted
as the spin-state crossover between the low-spin and high-spin state of $d^8$,
realized in our approximation by a level crossing. Using
Eq.~\eqref{eq:trialEavg}, the AF phase gives the average energy per site
$E_\mathrm{AF}=E_T-2J$, which has to be compared with $E_\mathrm{FQ}$ to find
the transition line observed in Fig.~\ref{fig:PD}(a). Following the AF/FQ
phase boundary from the large $t$ to small $t$ limit, we find the associated
step in the $\vc T$ density increasingly pronounced, reaching maximum at
$t=0$, where we hit the point corresponding to $E_T=0$. This point is a
trivially exact point of the phase diagram---here $E_T$ changes sign and in
the absence of exchange interactions due to $t=0$, the system immediately
switches between the state consisting exclusively of either $s$ or $\vc T$.

To assess the adequacy of the effective model and its simple variational
treatment, we performed exact diagonalization of the two-orbital Hubbard model
with \mbox{$e_g$-type} hopping on a small square-lattice cluster. The first
set of data to compare to the variational phase diagram is presented in
Fig.~\ref{fig:PD}(c). It shows the average triplet occupation $n_T$ per site
calculated for the 8-site cluster of Fig.~\ref{fig:PD}(d) by employing the
same microscopic parameters $\Delta$, $t$, $U$, and $J_\mathrm{H}$ as those
used to construct Fig.~\ref{fig:PD}(a). Note that, in contrast to $\rho$ of
Fig.~\ref{fig:PD}(a), $n_T$ of Fig.~\ref{fig:PD}(c) includes also the
fluctuating part of $\vc T$ density. While the map of $n_T$ itself gives
already a good idea of the phase diagram, to reliably identify the phases, we
check the tendency towards long-range order by calculating the characteristic
correlations $\langle\vc{S}\vc{S}\rangle_{\vc q=\vc Q}$ and
$\langle\vc{\widetilde{S}}\vc{\widetilde{S}}\rangle_{\vc q=\vc Q}$ 
associated with the AF and FQ phases, respectively. Shown in
Fig.~\ref{fig:PD}(e) are the ground-state values of the correlations evaluated
according to
$\langle\vc{S}\vc{S}\rangle_{\vc q}=N_\mathrm{site}^{-1}\sum_{j\neq j'}
\mathrm{e}^{i\vc q(\vc R_j-\vc R_{j'})}\langle\vc{S}_j\vc{S}_{j'}\rangle$,
i.e. excluding the on-site contribution. The overall agreement between
Fig.~\ref{fig:PD}(a) and Figs.~\ref{fig:PD}(c),(e) is more than satisfactory,
given that the variational phase diagram in fact involves two levels of
approximation: \textit{(i)} the effective model \eqref{eq:Hd8} was derived via
perturbation theory to second order in electron hopping, \textit{(ii)} the
effective model was solved using a crude variational Ansatz. Moreover, the
exact diagonalization for such a small cluster may be expected to overestimate
the effect of quantum fluctuations, shifting the phase boundaries
significantly. A comparison of the results for 4-, 6-, and 8-site cluster
indeed shows a systematic trend of the PM/FQ boundary moving slightly up with
increasing system size. Therefore, the Hubbard model calculation well
confirms the topology of the phase diagram and a rough location of the phase
boundaries observed in Fig.~\ref{fig:PD}(a), suggesting additionally a
smoother crossover between the high-spin AF phase and low-spin FQ phase.

Finally, Fig.~\ref{fig:PD}(f) demonstrates, that the overall structure of the
phase diagram does not change upon hole doping that we simulate by removing
one or two electrons from our Hubbard system, corresponding formally to 12.5\%
and 25\% doping. Though the data presented in Fig.~\ref{fig:PD}(f) are
necessarily plagued by large finite-size effects, we may anticipate roughly
the same location of the PM/FQ phase boundary and a growth of the FQ phase at
the expense of the AF phase. This trend is in line with intuition, since the
FQ phase characterized by the dynamical mixing of $s$ and $\vc T$ can better
accommodate the $d^7$ mobile carriers and their interplay with the $d^8$
background to be discussed in the next section. The FQ phase might be
therefore energetically favorable to the AF one upon doping. On the other
hand, from the same perspective there seems to be only a little difference 
between the FQ and PM phases.



\section{Doped case: $d^7$ holes within $d^8$ background}
\label{sec:d7}

In this section, the effective model will be completed by introducing mobile
carriers represented by holes corresponding to the $d^7$ configuration $f$
shown in Fig.~\ref{fig:ionic}(c). We only include the $d^7$ states with the
single $e_g$ electron occupying the $3z^2-r^2$ orbital. The participation of
$d^7$ configurations of $x^2-y^2$ character is partially suppressed by their
higher energy due to $\Delta$, partially it is caused by on-site correlations
preventing their motion even though the hopping amplitude associated with the
$x^2-y^2$ orbital is three times larger than that of $3z^2-r^2$. We will
discuss these issues at relevant points later. As a result, the $x^2-y^2$
electrons ``live'' in the system mostly bound in the triplet $d^8$
configurations.

In the following, we first discuss the derived interactions between the $d^7$
and $d^8$ objects in our model, then focus on the individual propagation of
the doped holes, and finally examine their pairing tendencies and implications
for possible superconductivity.

\subsection{Interaction between holes and background}
\label{sec:d7model}

\begin{figure}[tb]
\includegraphics[scale=1.0]{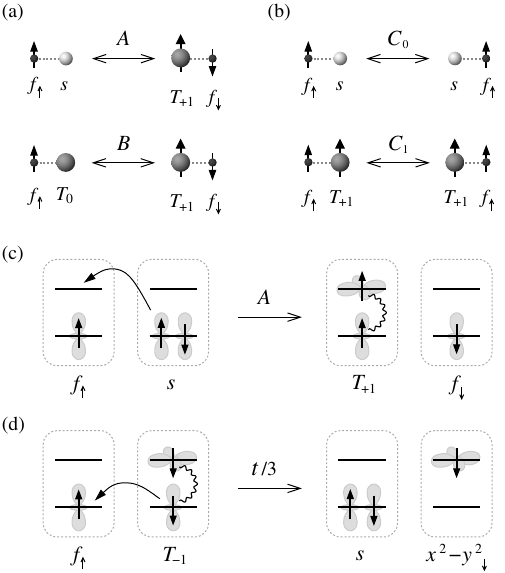}
\caption{
Cartoon representation of the $f$ particle motion through the $d^8$
background, i.e. examples of $d^7$--$d^8$ bond processes contained in
Eq.~\eqref{eq:Hd7d8}.
(a)~Hole hopping involving spin degrees of freedom: Within the process $A$,
the hole hopping is entangled with a singlet-triplet transition; during
process $B$, the hole exchanges its position with a triplon $\vc T$ in a
spin-sensitive manner.
(b)~Spin-independent hole hopping: the hole exchanges its position with either
a singlet $s$ (process $C_0$) or triplon $\vc T$ (process $C_1$) without
altering any spin state.
(c)~Process $A$ depicted using the representation of the spin-orbital states
as in Fig.~\ref{fig:ionic}(c). Its origin is the interorbital hopping of the
electron from $d^8$ to $d^7$ site. The electron added to the $x^2-y^2$
orbital gets bound into a triplon $\vc T$ by virtue of Hund's coupling.
(d)~Hopping process converting $3z^2-r^2$ hole $f_\uparrow$ into a virtual
$(x^2-y^2)_\downarrow$ fermion, at the energy cost of $3J_H$ required to break
$S=1$ triplet $T$ state at the neighboring site.
}\label{fig:d7d8model}
\end{figure}

The interaction of the $d^7$ holes with the $d^8$ background consisting of $s$
and $\vc T$ particles is derived in a straightforward way by projecting the electronic
hopping on a \mbox{$d^7$--$d^8$} bond onto the low-energy states constituting
the basis of our model. It takes a form of $f$ hopping accompanied by
``counterflow'' of $d^8$ objects $s$ and $\vc T$, which may involve changes in
their state or even an $s\leftrightarrow T$ transition. Several sample
processes of this kind are illustrated in Fig.~\ref{fig:d7d8model}.

To express the interaction in a compact form, we again utilize the bond $d^8$
operators $\vc{\widetilde{S}}_{ij}$ and $\vc{S}_{ij}$ defined by
Eq.~\eqref{eq:Sbonddef} and introduce new bond operators that capture the
hopping of $f$. The first one is a regular spin-independent hopping
\begin{equation}\label{eq:nijdef}
n_{ij}=f^\dagger_{i\uparrow}f^\pd_{j\uparrow}
+f^\dagger_{i\downarrow}f^\pd_{j\downarrow} \,,
\end{equation}
the second one involves the $f$ spin and is defined via
\begin{equation}\label{eq:sigmaijdef}
\sigma^\alpha_{ij} = \sum_{s,s'=\uparrow,\downarrow} 
f^\dagger_{is} \sigma^\alpha_{ss'} f^\pd_{js'}
\end{equation}
with $\sigma^\alpha$ ($\alpha=x,y,z$) denoting the Pauli matrices. In addition
to transferring $f$ between the sites, the latter operator performs either a
flip of the $f$ spin ($\sigma^x$, $\sigma^y$) or makes the hopping sensitive
to the spin value ($\sigma^z$). The final \mbox{$d^7$--$d^8$} Hamiltonian
reads as
\begin{multline}\label{eq:Hd7d8}
\mathcal{H}_{d^7\text{--}d^8} = \sum_{\langle ij \rangle}
\bigl[
\mathrm{sgn}_{ij}\, A\,\vc{\sigma}_{ij}\vc{\widetilde{S}}_{ji}
+B\,\vc{\sigma}_{ij}\vc{S}_{ji} 
\\
+n_{ij}(C_0 s_j^\dagger s_i^\pd + C_1 \vc{T}_j^\dagger \vc{T}_i^\pd)
\bigr]+\mathrm{H.c.}
\end{multline}
As in the case $\mathcal{H}_{d^8}$ of Eq.~\eqref{eq:Hd8}, the form of the
Hamiltonian $\mathcal{H}_{d^7\text{--}d^8}$ is again explicitly spin-isotropic
and in fact dictated by this symmetry. Analogous forms of the interaction
between holes and singlet-triplet background can be encountered e.g. in the
context of hole-doped Heisenberg ladders or bilayers \cite{Ede98,Voj99} or in
the context of spin-crossover cobaltates, where the triplons carry in addition
an orbital degree of freedom \cite{Cha07,Kha08}. Following the microscopic
derivation, the individual parameters entering \eqref{eq:Hd7d8} are given by
\begin{equation}\label{eq:d7d8pars}
A=\tfrac1{\sqrt6}t\cos\theta \,,\;
B=\tfrac12 t \,,\;
C_0=\tfrac13 t\cos^2\theta \,,\;
C_1=\tfrac12 t \,.
\end{equation}

A few examples of different types of processes contained in
$\mathcal{H}_{d^7\text{--}d^8}$ are sketched in
Fig.~\ref{fig:d7d8model}(a),(b). The most interesting one corresponds to the
$A$ term representing a hole hopping accompanied by $s\leftrightarrow T$
transition. Expanded in full, the relevant term
$\tfrac1{\sqrt2} \vc{\sigma}_{ij} \vc{\widetilde{S}}_{ji}$
takes the form
\begin{equation}
T^\dagger_{-1 j}\, f^\dagger_{i\uparrow} f^\pd_{j\downarrow}
+T^\dagger_{0 j}\, 
\tfrac1{\sqrt2}(f^\dagger_{i\uparrow} f^\pd_{j\uparrow}
 -f^\dagger_{i\downarrow} f^\pd_{j\downarrow})
-T^\dagger_{+1 j}\, f^\dagger_{i\downarrow} f^\pd_{j\uparrow}
+\text{H.c.}
\end{equation}
For brevity, we have omitted $s_i$ in each contribution and used the
$\tfrac1{\sqrt2}$ prefactor. The $A$ processes are essential to understand
the effects arising due to the hole doping in the PM phase of the background.
The dynamical generation and annihilation of triplet excitations $\vc T$ by
the doped hole has a profound impact on its propagation and leads to clear
polaronic features discussed in Sec.~\ref{sec:d7SCBA}. It may become also a
source of hole pairing that is evaluated in Sec.~\ref{sec:d7pairing}. When
focusing on the PM phase, the $A$ processes need to be considered together
with the processes linked to the parameter $C_0$, that just exchange the
position of $f$ and $s$ on the bonds and provide therefore a free motion of
$f$ through the predominantly $s$ background. The processes of corresponding
to the $B$ and $C_1$ terms become important at larger $\vc T$ density and are
thus relevant in case of the ordered $d^8$ background that is beyond our scope
here.

Another possibility of visualizing the hopping processes in
$\mathcal{H}_{d^7\text{--}d^8}$ ---based on the schematic representation of
the states of Fig.~\ref{fig:ionic}(c)---is used in
Fig.~\ref{fig:d7d8model}(c). Starting with the $f$ hole and its most frequent
bond-neighbor $s$ in the PM phase, the depicted interorbital hopping can be easily
associated with the $A$ process in Fig.~\ref{fig:d7d8model}(a).

We note that the hopping processes in the underlying $e_g$ orbital Hubbard
model allow a mixing of the $f$ hole of $3z^2-r^2$ symmetry with the
$d^7(x^2-y^2)$ states not included in our effective \mbox{$s$-$T$-$f$} model.
Figure~\ref{fig:d7d8model}(d) shows such an example, where an $f$-fermion gets
converted into a virtual $d^7(x^2-y^2)$ state. This process breaks $S=1$
triplet bound state and costs a large energy of $3J_H$. As the latter far
exceeds the matrix element $t/3$ of the $3z^2-r^2$ intraorbital hopping
involved in this process, the admixture of $x^2-y^2$ orbital into the
$f$-fermion wave function is small and can be neglected. This is confirmed
below by the ED analysis of the full $e_g$ orbital Hubbard model. Thus in our
effective \mbox{$s$-$T$-$f$} model, the $x^2-y^2$ electrons are present only
in a hidden form, i.e. bound in the triplon $\vc T$ states of $d^8$
configuration.


\subsection{Hole motion -- renormalization within SCBA}
\label{sec:d7SCBA}

In the previous section we found that the motion of doped holes proceeds via
rather complex bond processes, where the hopping of $d^7$ holes is
intertwined in various ways with triplet excitations in the $d^8$ background.
The aim of this section is to analyze, how the coupling to the triplet excitations
affects the motion of doped holes. We will focus specifically on the PM phase,
where it is sufficient to consider just $E_T$ and $\kappa$ terms from
$\mathcal{H}_{d^8}$ \eqref{eq:Hd8} describing the magnetic background and $A$
and $C_0$ processes from the $\mathcal{H}_{d^7\text{--}d^8}$ interaction
Hamiltonian \eqref{eq:Hd7d8}. The other terms that are higher-order in $\vc T$
operators can be omitted in the PM phase due to the low triplon density.

The bare propagation of the doped holes ($f$ particles) in the $d^8$
background, which is mostly composed of $s$ singlets in the PM phase, is
captured by the $C_0$ processes of \eqref{eq:Hd7d8} leading to the bare
dispersion $\varepsilon_{\vc k}=2C_0(\cos k_x+\cos k_y)$. The hopping
amplitude $C_0\approx \tfrac13 t$ [see Eq.~\eqref{eq:d7d8pars}] is derived
from the weak $3z^2-r^2$ intraorbital hopping, resulting in a small bare
bandwidth. The holes couple to the triplet excitations represented as before
by the triplons $T_\alpha$ ($\alpha=x,y,z$) of hardcore bosonic character,
carrying the elementary excitations of the $d^8$ background. The coupling is
provided by the $A$ term of \eqref{eq:Hd7d8}, which transforms into the
momentum representation as
\begin{equation}\label{eq:Atermk}
\mathcal{H}_{d^7\text{--}d^8} \rightarrow \;\;
i A \sum_{\vc k \vc q \alpha s s'} \eta_{\vc k-\vc q}
T^\dagger_{\alpha \vc q}\,
f^\dagger_{\vc k-\vc q,s} \sigma^\alpha_{ss'} f^\pd_{\vc k,s'} 
 \;+ \text{H.c.}
\end{equation}
with $\eta_{\vc q} = 2(\cos k_x-\cos k_y)$. The approximate diagonalization
of $\mathcal{H}_{d^8}$ as described in Sec.~\ref{sec:d8phases} leads to three
degenerate eigenmodes $\alpha_{\vc q}$ with the dispersion \eqref{eq:wqdisp}.
The corresponding operators are constructed from $T_{x,y,z}$ by Bogoliubov
transformation
$T_{\alpha\vc q}=u_{\vc q} \alpha^\pd_{\vc q} 
+ v_{\vc q} \alpha^\dagger_{-\vc q}$, 
where the Bogoliubov factors take the form
\begin{equation}
u_{\vc q} =\frac{1}{\sqrt2}\sqrt{ \frac{E_T+4\kappa\gamma_{\vc q}}{\omega_{\vc q}} + 1}
\,,\quad
v_{\vc q} =\frac{2\kappa\gamma_{\vc q}}{\omega_{\vc q} u_{\vc q}} \,.
\end{equation}
Inserted into Eq.~\eqref{eq:Atermk}, this gives the linear coupling of $f$ to
the excitations $\alpha_{\vc q}$:
\begin{equation}\label{eq:Atermkeig}
\mathcal{H}_{d^7\text{--}d^8} \rightarrow \;\;
i A \sum_{\vc k \vc q \alpha s s'} 
\Gamma_{\vc k\vc q} \alpha^\dagger_{\vc q} \,
f^\dagger_{\vc k-\vc q,s} \sigma^\alpha_{ss'} f^\pd_{\vc k,s'} 
\;+ \text{H.c.} \,,
\end{equation}
with the formfactor
$\Gamma_{\vc k\vc q} = \eta_{\vc k-\vc q} u_{\vc q} - \eta_{\vc k} v_{\vc q}$.
The propagation of $f$ through the $d^8$ background including the coupling
\eqref{eq:Atermkeig} to its excitations is treated within standard
selfconsistent Born approximation (SCBA) scheme commonly used to address
magnetic polarons (see e.g. \cite{Sch88,Kan89,Mar91,Nyh22}). For the case of a
single hole, we get the selfenergy
\begin{equation}
\Sigma(\vc k,E)=3A^2 \sum_{\vc q}
\Gamma_{\vc k\vc q}^2\,
\mathcal{G}(\vc k-\vc q,E-\omega_{\vc q}) \,,
\end{equation}
which enters the single-hole propagator 
$\mathcal{G}(\vc k,E) = [E+i0^+-\varepsilon_{\vc k}-\Sigma(\vc k,E)]^{-1}$ to
be determined selfconsistently.

\begin{figure}[tb]
\includegraphics[scale=1.0]{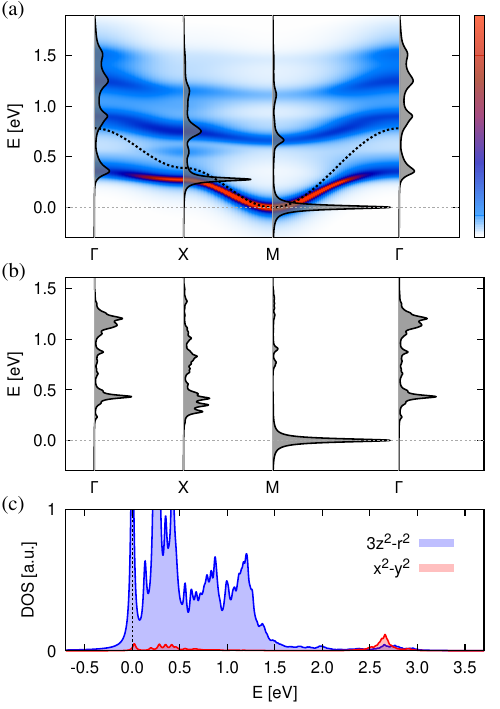}
\caption{
(a)~Single-hole spectral function obtained within SCBA applied to the effective
\mbox{$s$-$T$-$f$} model. The parameter point corresponds to
$\Delta=2\:\mathrm{eV}$ and $t=0.3\:\mathrm{eV}$ in Fig.~\ref{fig:PD}(a).
The spectral function is presented both in a form of a color map as well as
profiles at selected high-symmetry points in the Brillouin zone. The dotted 
line shows the bare dispersion $\varepsilon_{\vc k}$. 
(b)~Spectral function of a $3z^2-r^2$ hole obtained by exact diagonalization
of the $e_g$ Hubbard model at the 8-site cluster for the same parameter point. 
Both spectral functions were artificially broadened for an easier comparison
and shifted to measure the energy from the bottom of the band.
(c)~Single-particle density of states (DOS) for the 8-site cluster resolved
according to the contributing $e_g$ orbital. The DOS corresponds to an average
of the spectral function over the eight accessible momenta in the Brillouin zone.
}\label{fig:SCBA}
\end{figure}

Presented in Fig.~\ref{fig:SCBA}(a) is the corresponding spectral function
$A(\vc k,E)=-\pi^{-1}\mathrm{Im}\,{\mathcal{G}(\vc k,E)}$ obtained for a
representative parameter point that falls into the PM phase. The spectral
function bears typical polaronic features. The bottom of the hole band around
$\vc k=M$ is only mildly affected, but the higher parts are strongly
renormalized with the quasiparticle weight reduced by factor $\approx 2-3$.
Compared to the bare dispersion, the total bandwidth is about two times
smaller. The energy position of the onset of the damping and dispersion
flattening correlates with $E_T$, in this case evaluated to $\approx
0.29\:\mathrm{eV}$.

The spectral function resulting from the effective model is well confirmed by
exact diagonalization of the underlying Hubbard model. In
Fig.~\ref{fig:SCBA}(b) we present the spectral function corresponding to an
extraction of \mbox{$3z^2-r^2$} electron
\begin{equation}
A_{\mathrm{ED}}(\vc k, E) =
-\frac{1}{\pi}\,\mathrm{Im}\,\langle \mathrm{GS} | z^\dagger_{\vc k\sigma}
\;\frac{1}{E+i0^{+}-\mathcal{H}_\mathrm{Hubb}}\; z_{\vc k\sigma} |\mathrm{GS} \rangle\,,
\end{equation}
which may be directly compared to the $f$ spectral function. The calculation
was performed for the 8-site cluster of Fig.~\ref{fig:PD}(d) used before.
$|\mathrm{GS}\rangle$ denotes the cluster ground state. Given the smallness
of the cluster and the approximations made in the derivation and treatment of
the effective model, the agreement is very good. This supports the picture
that we have drawn based on the effective model, i.e. hole propagation along
with intense generation and annihilation of the triplet excitations.

In Fig.~\ref{fig:SCBA}(c) we return to the question of the relevance of the
$x^2-y^2$ holes for the effective model, as discussed in the end of
Sec.~\ref{sec:d7model}. Using the exact diagonalization of the Hubbard model,
we can directly quantify the degree of their involvement. To this end, we
study the orbital-resolved single-particle density of states, evaluated as an
average of the corresponding spectral function over the cluster-compatible
momenta $\vc k$: $\mathrm{DOS}(E) = N_{\vc k}^{-1} \sum_{\vc k} A(\vc k,E)$.
Here $A(\vc k,E)$ is either the spectral function for the $3z^2-r^2$ hole
presented for selected $\vc k$ points in Fig.~\ref{fig:SCBA}(b), or analogous
one for the $x^2-y^2$ hole calculated as 
$-\pi^{-1}\,\mathrm{Im}\,\langle \mathrm{GS} | x^\dagger_{\vc k\sigma}
(E+i0^{+}-\mathcal{H}_\mathrm{Hubb})^{-1} x_{\vc k\sigma} |\mathrm{GS} \rangle$.
As seen in Fig.~\ref{fig:SCBA}(c), the DOS related to the $x^2-y^2$ hole is
indeed negligible compared to the $3z^2-r^2$ one and is concentrated in two
energy regions. The first one below approximately $0.5\:\mathrm{eV}$ covers
essentially the range of the low-energy band of propagating $3z^2-r^2$ hole.
This feature indicates a slight mixing of the $x^2-y^2$ and $3z^2-r^2$ holes
stemming from the hole-type conversion processes discussed in the context of
Fig.~\ref{fig:d7d8model}(d). The second region is located at energies higher
by about $\Delta$, which corresponds to the expected excitation energy of the
$x^2-y^2$ hole as compared to the $3z^2-r^2$ one.


\subsection{Hole pairing and BCS estimates}
\label{sec:d7pairing}

Focusing on the PM phase again, we now consider the dynamic generation and
annihilation of triplet excitations as a source of pairing among the holes.
We will follow the usual procedures applied in the theory of conventional
superconductivity based on electron-phonon coupling and derive the hole-hole
interaction by perturbatively eliminating the coupling Hamiltonian
\eqref{eq:Hd7d8} that connects the $d^7$ and $d^8$ sectors, obtaining thereby
the effective interaction acting in the $d^7$ sector.

Specifically, from $\mathcal{H}_{d^7\text{--}d^8}$ of Eq.~\eqref{eq:Hd7d8} we
again take the $A$-channel part, here in the form
$\sum_{ij}\mathrm{sgn}_{ij}\,A\,\vc{\sigma}_{ij}\vc{\widetilde{S}}_{ji}$.
The sum now runs through all nearest-neighbor $i$, $j$ and includes thus also
the H.c. counterpart of the bond sum in \eqref{eq:Hd7d8}. The $d^7$ hole
operators are embedded in the bond operators $\vc{\sigma}_{ij}$ with the
components given by \eqref{eq:sigmaijdef}, while the $d^8$ sector is
represented by $\vc{\widetilde{S}}_{ji}$ defined by \eqref{eq:Sbonddef}. By
eliminating the above $d^7$--$d^8$ coupling via second-order perturbation
theory, we obtain the effective interaction among $f$, which can be expressed
in the form
\begin{equation}\label{eq:Hff}
\mathcal{H}_{f\text{--}f} \approx 
-A^2 \sum_{ij \alpha} \sum_{kl \beta} \mathrm{sgn}_{ij}\,\mathrm{sgn}_{kl}\;
F^{\alpha\beta}_{ji,lk}\; \sigma^\alpha_{ij}\,\sigma^\beta_{kl}
\end{equation}
with $F$ being the $\langle \widetilde{S}\widetilde{S}\rangle$ propagator
evaluated entirely within the $d^8$ sector. It is defined as the $d^8$
ground-state average
\begin{equation}\label{eq:Fdef}
F^{\alpha\beta}_{ji,lk} = 
\Bigl\langle \mathrm{GS} \,\Bigl |
\;{\widetilde{S}}^\alpha_{ji}\;
\frac{1}{\mathcal{H}_{d^8}-E_\mathrm{GS}}
\;{\widetilde{S}}^\beta_{lk}\;
\Bigl |\, \mathrm{GS} \Bigr\rangle \,.
\end{equation}
Recalling the similarity to the electron-phonon problem, the propagator $F$
entering the hole-hole interaction is analogous to the static limit of the
phonon propagator mediating effective electron-electron interaction. Thanks
to the spin-space symmetry of the $d^8$ background, the quantity
\eqref{eq:Fdef} is isotropic,
$F^{\alpha\beta}_{ji,lk}=F_{ji,lk}\delta_{\alpha\beta}$, and we need to
consider just the diagonal elements $F_{ji,lk}$ in the following. Note that
by construction, the pairs of indices $i$, $j$ and $k$, $l$ correspond to
nearest neighbors (NN) and we thus deal with a bond-bond correlator.

\begin{figure}[t!b]
\includegraphics[scale=1.0]{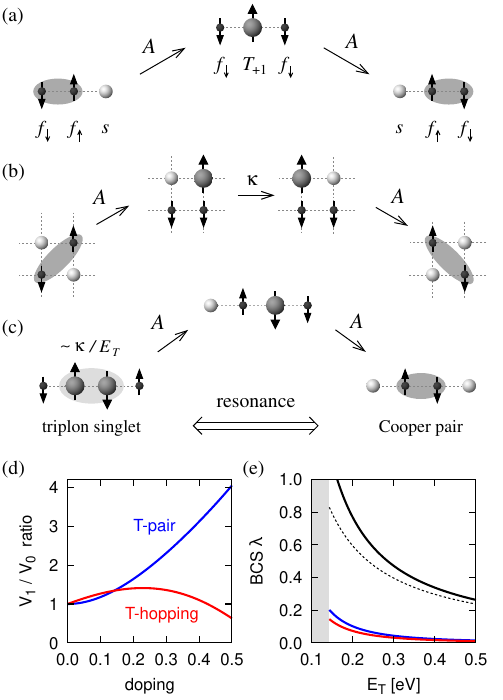}
\caption{
(a)~Cartoon representation of the dominant pairing process taking place on two
adjacent bonds. One of the $f$ holes undergoes $A$ hopping leaving a triplon
$\vc T$ excitation behind, which is later absorbed by the other hole in the 
second $A$ hopping. The amplitude of this process is proportional to $A^2/E_T$.
(b),(c)~Examples of additional pairing processes arising as corrections to (a)
in the first order in $\kappa/E_T$. Process (b) is similar to (a) but with an
extra hopping $\kappa$ of the intermediate triplon $\vc T$.
The process (c) corresponds to an absorption of a singlet pair of triplons
$\vc T$ (created via the $\kappa$ exchange channel and having an amplitude
$\propto \kappa/E_T$) by a pair of $f$ fermions. The absorption occurs via two
$A$ hoppings with the intermediate state containing one unpaired $\vc T$,
hence its amplitude is $\propto A^2/E_T$. Such a hopping chain may proceed in
both directions, leading to a promotion of $f$ singlets by their resonance
with the singlet triplon pairs fluctuating in the $d^8$ background by virtue
of the $\kappa$ exchange.
(d)~Relative contribution of first-order terms to $s$-wave pairing potential
\eqref{eq:VKK} averaged over a circular Fermi surface at various doping
levels. Following Eq. \eqref{eq:VKK}, the numbers have to be multiplied by
$\kappa/E_T$ when comparing the absolute pairing strength.
(e)~BCS $\lambda$ for fixed $t=0.3\:\mathrm{eV}$ and varying $\Delta$
presented as function of $E_T$ at doping level $0.25$. Total value is shown as
black solid line, contributions from $V_0$, $\vc T$-hopping, and $\vc T$-pair
first-order terms are shown as dashed/red/blue lines, respectively. $E_T$
interval corresponding to long-range order is indicated by shading.
}\label{fig:BCS}
\end{figure}

The evaluation of $F_{ji,lk}$ for the PM phase is performed using an
approximate solution of the $d^8$ model limited to $E_T$ and $\kappa$ terms as
in Sec.~\ref{sec:d7SCBA}. One can proceed, for example, by converting
\eqref{eq:Fdef} to the momentum representation and utilizing the previous
description of the PM phase based on the Bogoliubov transformation of $\vc T$.
For simplicity and to enable a transparent interpretation of the result, we
will limit ourselves to an expansion to first order in $\kappa/E_T$, where
$F_{ji,lk}$ can be also easily evaluated using real-space perturbation theory
and the contributions visualized in a pictorial way as done in
Fig.~\ref{fig:BCS}(a)-(c). In this expansion, we start with the $E_T$ part of
$\mathcal{H}_{d^8}$ of \eqref{eq:Hd8} as the unperturbed Hamiltonian and add
the \mbox{$\kappa$-interaction} playing the role of the perturbation. Up to
first order in $\kappa/E_T$, we get
\begin{equation}\label{eq:Fjilkeval}
F_{ji,lk} = \frac1{E_T}\left[ \delta_{il} - \frac{\kappa}{E_T}
(\delta_{il}^\mathrm{NN}+\tfrac12\delta_{ik}^\mathrm{NN}
+\tfrac12\delta_{jl}^\mathrm{NN})
\right].
\end{equation}

The leading term of zeroth order contains standard Kronecker $\delta_{il}$
which is equal to unity if $i=l$. The physical process behind this
contribution to the effective \mbox{$f$--$f$} interaction \eqref{eq:Hff} is a
creation of triplon by hopping of $f$ during the first $A$ process and its
immediate absorption in the second $A$ process [see Fig.~\ref{fig:BCS}(a) for
a cartoon representation]. The possibility of such a pair hopping and the
related kinetic energy gain drives the pairing of the holes into short-range
pairs. Higher-order processes, that involve some action of $\kappa$, give
rise to a more extended attraction of holes. In the first order included in
\eqref{eq:Fjilkeval}, there are two physically distinct contributions that can
be associated with the processes depicted in Fig.~\ref{fig:BCS}(b),(c). The
first contribution corresponds to the term $\delta^\mathrm{NN}_{il}$ that
checks whether the sites $i$ and $l$ form a nearest-neighbor bond. The process
behind is similar to that of Fig.~\ref{fig:BCS}(a), but there is an extra
hopping of the triplon between sites $i$ and $l$ in the intermediate state,
which is provided by the $\kappa$ interaction in the $d^8$ sector. The second
contribution, linked to the combination
$\frac12(\delta^\mathrm{NN}_{ik}+\delta^\mathrm{NN}_{jl})$ 
in \eqref{eq:Fjilkeval}, has a completely different physical origin. It
exploits the fact, that the $\kappa$ interaction preforms singlet pairs of
triplons $\vc T$, which may---through two correlated hoppings within the
$A$-channel of $\mathcal{H}_{d^7\text{--}d^8}$ ---resonate with singlet pairs
of holes. The triplon pairs appear with amplitude $\propto\kappa/E_T$, hence
such processes appear as first-order in \eqref{eq:Fjilkeval}.

The next step is to insert the approximate $F_{ji,lk}$ into \eqref{eq:Hff} and
collect all the possible process pathways. Here the real-space formulation
proves quite helpful, as it enables to handle the various constraints
explicitly. The result is converted into momentum space and adapted to the
BCS form
\begin{equation}
\mathcal{H}_\mathrm{BCS} \approx \sum_{\vc k\vc k'} V_{\vc k\vc k'}
f^\dagger_{\vc k\uparrow} f^\dagger_{-\vc k\downarrow}
f^\pd_{-\vc k'\downarrow} f^\pd_{\vc k'\uparrow} \,.
\end{equation} 
We have only picked up hole pairs with zero total momentum and the singlet
pairing channel (triplet hole pairs come with an overall reduction by a factor
of $3$). The matrix elements $V_{\vc k\vc k'}$ have somewhat complicated form,
as the interaction reaches beyond NN due to the terms $\propto \kappa/E_T$.
For convenience, we introduce two sets of symmetry-adapted functions used to
capture the momentum dependence of pairing interaction in \mbox{$s$-wave} and
\mbox{$d$-wave} channel. The \mbox{$s$-wave} symmetry functions read as 
\begin{align}
s_1 &=\cos k_x + \cos k_y \,, \notag \\
s_2 &=\cos 2k_x + \cos 2k_y \,, \notag \\
s_3 &=\cos 3k_x + \cos 3k_y \,, \notag \\
s_4 &=\cos k_x\cos 2k_y + \cos 2k_x\cos k_y \,, \notag \\
s_5 &=\cos k_x \cos k_y \,, \label{eq:swaveset}
\end{align}
while for the \mbox{$d$-wave} symmetry we define the following ones
\begin{align}
d_1 &=\cos k_x - \cos k_y \,, \notag \\
d_2 &=\cos 2k_x - \cos 2k_y \,, \notag \\
d_3 &=\cos 3k_x - \cos 3k_y \,, \notag \\
d_4 &=\cos k_x\cos 2k_y - \cos 2k_x\cos k_y \,. \label{eq:dwaveset}
\end{align}
The matrix elements entering the pairing interaction are expressed following
the structure of \eqref{eq:Fjilkeval}:
\begin{equation}\label{eq:VKK}
V_{\vc k\vc k'} = \frac{3A^2}{E_T} \left[ V_{0\vc k\vc k'}
+\frac{\kappa}{E_T}\, V_{1\vc k\vc k'} + 
\mathcal{O}\left(\frac{\kappa^2}{E_T^2}\right) \right] \,.
\end{equation}
In units of $3A^2/E_T$, the $s$-channel contribution of the zeroth and first
order in $\kappa/E_T$, stemming from $F_{ji,lk}$ of Eq.~\eqref{eq:Fjilkeval},
reads as
\begin{multline}
V^{(s)}_{\vc k\vc k'} \approx
-2s_1 s'_1 
+\frac{\kappa}{E_T} \Bigl[
-32 s_5 s'_5 + 8 (s_2 s'_5 + s_5 s'_2) - 2 s_2 s'_2 \\
 -4 s_1 s'_1 + 2(s_1 s'_4 + s_4 s'_1)-(s_1 s'_3 + s_3 s'_1) \Bigr] \,.
\end{multline}
Here, the shorthand notations $s_n$ and $s'_n$ imply the corresponding
functions from the set \eqref{eq:swaveset} with the argument $\vc k$ or 
$\vc k'$, respectively. The first-order term ($\propto \kappa/E_T$) in the 
first line comes from \mbox{$\vc T$-hopping} in the virtual state, the second 
line is due to the resonance between $f$-pairs and $\vc T$-pairs. Similarly, 
the \mbox{$d$-wave} contribution can be written as
\begin{multline}
V^{(d)}_{\vc k\vc k'} \approx
+6 d_1 d'_1 
+\frac{\kappa}{E_T} \Bigl[
-2 d_2 d'_2 \\
 -4 d_1 d'_1 -6(d_1 d'_4 + d_4 d'_1)-(d_1 d'_3 + d_3 d'_1) \Bigr] \,.
\end{multline}

After specifying the pairing interaction, we now apply it in a situation with
moderate hole doping \mbox{$\lesssim 0.3$}, where the $d^8$ background is
still expected to retain its specific features. The dispersion of the hole
band is roughly parabolic with a minimum at \mbox{$\vc k=M$} point. Under
such conditions, the leading term in the $d$-wave channel is obviously
disfavored, we thus focus solely on the $s$-wave channel. Assuming nearly
constant superconducting gap, we average the pairing potentials over
\mbox{$M$-centered} circular Fermi surface according to the formula
$\langle V\rangle_\mathrm{FS} = 1/(2\pi k_\mathrm{F})^2
\oint_\mathrm{FS} \mathrm{d}\vc k
\oint_\mathrm{FS} \mathrm{d}\vc k'
\;V_{\vc k\vc k'}$, which is 
evaluated at various doping levels $x$ entering through the Fermi radius
$k_\mathrm{F}=\sqrt{2\pi x}$. In Fig.~\ref{fig:BCS}(d) we first compare the
relative contributions of the individual types of processes shown in
Fig.~\ref{fig:BCS}(a)-(c) to the average 
$\langle\,V^{(s)}_{\vc{k}\vc{k'}}\,\rangle_\mathrm{FS}=V_0+({\kappa}/{E_T})V_1$.
Relative to the leading term $V_0$ associated with the process in
Fig.~\ref{fig:BCS}(a), the total first order contribution
[Fig.~\ref{fig:BCS}(b),(c) taken together] comes with a factor of about $3-4$
multiplied additionally by $\kappa/E_T$. Considering the limitation of the
latter ratio by the critical value $\kappa/E_T=1/8$, the first-order
contribution does not exceed about $30\%$ of the total pairing strength in
this approximation. Although this leaves the zeroth order dominant, $T_c$ may
be enhanced significantly by the two kinds of first-order processes. Among
them, the pair resonance [Fig.~\ref{fig:BCS}(c)] takes over at larger doping
levels.

We are now ready to make a rough BCS estimate. Plotted in
Fig.~\ref{fig:BCS}(e) is the averaged pairing potential multiplied by the bare
density of states $1/4\pi C_0$, forming together the BCS parameter
$\lambda$:
\begin{equation}
\lambda = -\,\langle\, V^{(s)}_{\vc k\vc k'}\, \rangle_\mathrm{FS} 
\; \frac{1}{4\pi C_0} \,.
\end{equation}
Here, $C_0$ is determined by hopping $t$, see Eq.~\eqref{eq:d7d8pars}. The
data in Fig.~\ref{fig:BCS}(e) correspond to a representative
$t=0.3\:\mathrm{eV}$ cut through the phase diagram of Fig.~\ref{fig:PD}(a). By
varying $\Delta$ as in Fig.~\ref{fig:PD}(a), we mainly tune the parameter
$E_T$. Its values are used in Fig.~\ref{fig:BCS}(e) instead of $\Delta$ to
produce a more transparent horizontal axis. As seen in Fig.~\ref{fig:BCS}(e),
in the regime not very close to the PM/FQ phase boundary, the first-order
contributions represent about $20\%$ of the total $\lambda$. One can thus
expect the higher-order terms in the pairing potential \eqref{eq:VKK} to have
a relatively minor effect. To make a conservative estimate, we take
$\lambda\approx 0.5$ for $E_T=0.3\:\mathrm{eV}$ and insert it into the BCS
formula $T_c\approx 1.13\,\Omega\,\exp(-1/\lambda)$. The cutoff $\Omega$ is
limited by the $E_T$ value $\approx 0.3\:\mathrm{eV}$ but also by the quite
small bandwidth. By setting for example $\Omega\approx 0.1\:\mathrm{eV}$, we
would arrive at $T_c$ of about $15\:\mathrm{meV}\approx 180\:\mathrm{K}$. Even
though such estimate is of course naive, the values of $\lambda$ are promising
for a potential superconductivity in nickel-based compounds displaying the
spin-state crossover phenomenon. This is due to a rather strong coupling to
triplet excitations with relatively high energy, combined with the enhanced
density of states for a narrow band derived from the $3z^2-r^2$ orbital.



\section{Conclusions}
\label{sec:Conclusions}

We presented a basic theory for hole-doped nickelates containing low-spin
$d^8$ Ni$^{2+}$ ions on a square lattice. Such an ionic ground state may be
stabilized by various means, e.g., by square-planar coordination of Ni ions,
substitution of apical oxygens by halide ions, or by an extreme elongation of
the \mbox{NiO$_6$} octahedra in the out-of-plane direction. We specifically
focused on the spin-crossover regime, where---by a proper balance of the
crystal field splitting and Hund's coupling---the low-spin $S=0$ ionic state
is quasidegenerate with the high-spin $S=1$ state. In this regime, the
material would be described by a rich magnetic model of singlet-triplet type,
that we have derived perturbatively and supported by exact diagonalization of
the underlying Hubbard model based on $e_g$ orbitals. 
After establishing the relevant phase diagram and excitation spectrum, we
considered a doped material containing a moderate proportion of $d^7$
Ni$^{3+}$ ions. By electronic hopping, these $d^7$ configurations act as
mobile hole-like carriers that show a non-trivial interplay with the $d^8$
background that we studied following again a microscopically derived model.
The propagation of the doped holes is found to be tightly intertwined with a
dynamic creation and annihilation of triplet excitations (triplons) in the
$d^8$ background, leading e.g. to strong polaronic features in their spectral
function.
Most importantly, the processes involving triplet excitations generate an
effective interaction among the holes, which provides a strong Cooper pairing,
suggesting $s$-wave superconductivity in the spin-crossover materials. Part
of this pairing potential is contributed by an interesting new mechanism based
on a resonance between the singlet pairs of doped holes and the singlet pairs
of charge-neutral triplons created in the $d^8$ background by virtue of the
exchange interactions. This mechanism is generic to doped singlet-triplet
models and though it brings a relatively small contribution in the studied
case, it may be strong in situations with high densities of triplons entangled
in total-singlet pairs, such as that encountered in the highly frustrated
Kitaev-like singlet-triplet model developed in the context of $d^4$ honeycomb
ruthenates \cite{Ani19,Cha19}. In general, the exploration of the
unconventional superconductivity in the spin-state crossover materials,
especially in low-spin Ni$^{2+}$ $d^8$ compounds is an attractive direction
for future theoretical and experimental studies.

Finally, let us note that while we considered a homogeneous state upon doping,
spatial inhomogeneities may develop in certain situations.
The presence of doped carriers generally brings frustration to the magnetic
interactions of the background, which might be relieved for example by phase
separation or stripe formation, as reported for the related two-orbital
Hubbard models (see e.g. \cite{Sbo07,Rac06}).
In the regime of our primary focus, i.e. the paramagnetic phase supporting
singlet superconductivity, such tendencies are not expected as the triplon
density is low and a homogeneous singlet background is stabilized by a large
energy cost of a triplon creation. However, once the triplons are at low
energies, dense and interacting intensely, the doped holes introduced into
such system may lead to a formation of a certain pattern, such as stripes,
possibly supported by the hole-triplon interactions we obtained. Though our
numerical treatment of the Hubbard model was not able to suggest such a
possibility due to the very limited cluster size, this is certainly an
interesting follow-up problem and becomes highly relevant for nickelate
materials on the verge between the low-spin and high-spin states.
In a broader context, another intriguing problem is the evolution of the
system captured by $e_g$ Hubbard model as function of the crystal-field
splitting $\Delta$. In the parameter regime considered in the present work,
the effective singlet-triplet model proved quite robust, being stabilized by
large $\Delta$, and a reduction of $\Delta$ merely translates to a decrease of
singlet-triplet splitting, giving potentially rise to the inhomogeneities
mentioned above. Decreasing $\Delta$ further, the $x^2-y^2$ electrons will get
released from the triplon bound states and one should recover the usual
two-band Hubbard model physics. The details of this transition are a
challenging open question.


\acknowledgments

We would like to thank M.~Hepting and P.~Puphal for fruitful discussions
concerning the material aspects.
\mbox{J. Ch.} was supported by the Czech Science Foundation (GA\v{C}R) under
Project No.~\mbox{GA22-28797S} and by the project Quantum Materials for
Applications in Sustainable Technologies, Grant
No.~CZ.02.01.01/00/22\_008/0004572. 
\mbox{G. Kh.} acknowledges support from the European Research Council under
Advanced Grant no. 101141844 (SpecTera). 
Computational resources were provided by the e-INFRA CZ project (ID:90254),
supported by the Ministry of Education, Youth and Sports of the Czech
Republic.


\section*{Data availability}

The data that support the findings of this article are openly available
\cite{repository}.


\bibliography{paper}

\end{document}